# AI Explainability and Governance in Smart Energy Systems: A Review

**Roba Alsaigh[1], Rashid Mehmood[2]\*, Iyad Katib[1]**

[1]Department of Computer Science, Faculty of Computing and Information Technology (FCIT), King Abdulaziz University, Jeddah 21589, Saudi Arabia

[2]High Performance Computing Center, King Abdulaziz University, Jeddah 21589, Saudi Arabia

**\* Correspondence:**
Corresponding Author
RMehmood@kau.edu.sa

**Abstract**

Traditional electrical power grids have long suffered from operational unreliability, instability, inflexibility, and inefficiency. Smart grids (or smart energy systems) continue to transform the energy sector with emerging technologies, renewable energy sources, and other trends. Artificial intelligence (AI) is being applied to smart energy systems to process massive and complex data in this sector and make smart and timely decisions. However, the lack of explainability and governability of AI is a major concern for stakeholders hindering a fast uptake of AI in the energy sector. This paper provides a review of AI explainability and governance in smart energy systems. We collect 3,568 relevant papers from the Scopus database, automatically discover 15 parameters or themes for AI governance in energy and elaborate the research landscape by reviewing over 150 papers and providing temporal progressions of the research. The methodology for discovering parameters or themes is based on "deep journalism", our data-driven deep learning-based big data analytics approach to automatically discover and analyse cross-sectional multi-perspective information to enable better decision-making and develop better instruments for governance. The findings show that research on AI explainability in energy systems is segmented and narrowly focussed on a few AI traits and energy system problems. This paper deepens our knowledge of AI governance in energy and is expected to help governments, industry, academics, energy prosumers, and other stakeholders to understand the landscape of AI in the energy sector, leading to better design, operations, utilisation, and risk management of energy systems.

## 1    Introduction

Energy has fundamentally shaped the geopolitics of our world and transformed our lives in the last century (Vakulchuk, Overland and Scholten, 2020; Blondeel *et al.*, 2021). A look at the global past and current conflicts reveal that energy has been central to many of them involving oil, natural gas, battery minerals, among others. Energy availability enabled modern technological advancements including the ubiquity of computing and power (e.g., batteries), and transformed us into smart societies.

Traditional electrical power grids have long suffered from operational unreliability, instability, inflexibility, and inefficiency. Since power systems traditionally comprised large regional and national grids, monitoring the electrical systems of those grids and long distribution lines has been challenging causing many major electrical system failures, human lives, and hefty economic losses.

Smart grids continue to transform the energy sector with emerging technologies, renewable energy sources, decentralisation, decarbonisation, and others. We hereon will use the term "smart energy systems" for "smart grid" as a broader term that incorporates smart grids, electrical power systems, and related business and other developments. These advancements offer many exciting opportunities such as the availability of solar, wind, hydro, and other forms of energy to organisations and homes. Development of microgrids (Hussain, Bui and Kim, 2019), mini-grids (Gill-Wiehl et al., 2022), community grids (Ceglia et al., 2020; Kong and Song, 2020), and supergrids (Zarazua de Rubens and Noel, 2019) have paved the way, alongside many other possibilities, for energy independence and energy trading between individuals, corporations, and nations. These smart energy systems are complex and produce massive data.

AI presents an unimaginable potential for innovation, process optimisation, productivity, and other benefits in many sectors such as smart societies (Janbi et al., 2022), healthcare (Alahmari et al., 2022), education (Mehmood et al., 2017), and transportation (Alomari et al., 2021). The energy sector is not an exception (Alkhayat, Hasan and Mehmood, 2022). Artificial intelligence (AI) is being applied to smart energy systems to process massive and complex data in the energy sector and make smart and timely decisions. AI algorithms are black-box (Castelvecchi, 2016) needing interpretability and explainability (Doran, Schulz and Besold, 2017; Goebel et al., 2018; Hagras, 2018; Hoffman et al., 2018; Lundberg et al., 2020) so that the decision made by AI could be explained to various stakeholders such as for regulatory and legal reasons. AI algorithms are usually imperfect or inaccurate. These AI algorithms are developed by human designers and developers trained using imperfect data and therefore they are likely to inherit bias and prejudice from them. The unregulated developments of AI have focussed on maximising efficiencies, and economic and other objectives rather than human values and priorities.

We adopt in this paper the definition of explainable AI by NIST (National Institute of Standards and Technology) (Phillips et al., 2021) that proposes explainable AI systems to observe four principles, namely (i) Explanation ("a system delivers or contains accompanying evidence or reason(s) for outputs and/or processes"); (ii) Meaningful ("a system provides explanations that are understandable to the intended consumer(s)"); (iii) Explanation Accuracy ("an explanation correctly reflects the reason for generating the output and/or accurately reflects the system's process"); and (iv) Knowledge Limits ("a system only operates under conditions for which it was designed and when it reaches sufficient confidence in its output"). Explainability and interpretability are among many desirable characteristics to support trustworthiness in AI systems including, among others, "accuracy, privacy, reliability, robustness, safety, security (resilience), mitigation of harmful bias, transparency, fairness, and accountability" (Phillips et al., 2021). Responsibility could be another characteristic for AI trustworthiness (Yigitcanlar et al., 2021).

The lack of explainability and governability of AI had affected stakeholders' confidence in AI systems and consequently, the uptake of AI in the energy sector has been slow. Moreover, the complexity of the design and operations space of energy systems that involves many parameters and stakeholders is on the rise and the consequent severity of risks is catastrophic due to the social, national, environmental, and geopolitical criticality of these matters.

This paper provides a review of AI explainability and governance in smart energy systems. We collected 3,568 relevant papers from the Scopus database using a specific query (see Section 2), automatically discovered 15 parameters for AI governance in smart energy systems, and group them into four macro-parameters, namely AI Behaviour & Governance, Technology, Design & Development, and Operations. We elaborate on the research landscape by reviewing over 150 papers and providing temporal progressions of the research. The methodology for discovering parameters or themes is based on "deep journalism", our data-driven deep learning-based big data analytics approach to automatically discover and analyse cross-sectional multi-perspective information to enable better decision-making and develop better instruments for governance. We introduced the deep journalism approach (Ahmad *et al.*, 2022) and applied it to different sectors (Alqahtani *et al.*, 2022; Alswedani *et al.*, 2022; Mehmood, 2022).

The findings of this paper show that research on AI explainability in energy systems is segmented and narrowly focussed on a few AI traits (fairness, interpretability, explainability, trustworthiness) and energy system problems (stability and reliability analysis, energy forecasting, power system flexibility). The paper deepens our knowledge of AI governance in energy and is expected to help governments, industry, academics, energy prosumers, and other stakeholders to understand the landscape of AI in the energy sector, leading to better design, operations, utilisation, and risk management of energy systems.

## 1.1 Related Works and Novelty

To the best of our knowledge, this is the first comprehensive review of AI governance in the energy sector. It is a novel work due to its scope, methodology, and findings. There are several literature reviews on smart grids but they are not aimed at AI explainability or governance. We have found only two works that can be considered related to our work. A review of AI interpretability in smart grids is presented by (Xu *et al.*, 2022) using papers collected from Google Scholar over a five-year period. A review of AI explainability research in energy and power systems is provided by (Machlev *et al.*, 2022) using literature from 2019 to 2022. Firstly, none of these works have used BERT (NLP) to automatically discover parameters. Secondly, they do not have similar scope to ours (search query and data collection), i.e., they do not consider AI explainability and governance in its broader sense incorporating AI behaviour (governance, explainability, interpretability, responsibility, ethics, trustworthiness, and fairness) as extensively as we do (see Section 2). Thirdly, our deep journalism methodology allows us to use AI to collect a comprehensive selection of papers (a dataset) and provide a summary of a 55-year period of research on AI in energy systems.

The rest of the paper is organised as follows. Section 2 briefly describes the methodology of this work. Section 3 discusses the parameters and reviews the literature. Section 4 provides a discussion and concludes the paper.

## 2    Methodology and Design

We briefly describe the methodology and software tool design for automatic parameter discovery here. The word limit limits us, hence the brevity, for details, see (Ahmad *et al.*, 2022; Alqahtani *et al.*, 2022).

We collected the data for this work from the Scopus database using the following keywords in the query: artificial intelligence, machine learning, deep learning, grid, electricity, energy, power system, governance, explainability, interpretability, responsibility, ethics, trustworthiness, and fairness. This generated 3,568 paper abstracts published between 1967 and 2022 from various disciplines. No limits on disciplines or years were applied in collecting data. Duplicates, stop words, and irrelevant and noisy data were removed using pandas and NumPy. BERT (bidirectional encoder representations from transformers), UMAP (uniform manifold approximation and projection), HDBSCAN (hierarchical density-based spatial clustering of applications with Noise), and class-based TF–IDF (term frequency-inverse document frequency) score were used to capture contextual relations, reduce the number of clusters, and cluster data (Grootendorst, 2021; Ahmad *et al.*, 2022; Alqahtani *et al.*, 2022). Finally, we used domain knowledge and a range of analysis and visualisation techniques (hierarchical clustering, topic word score, similarity matrix, term score decline) to discover parameters for AI governance in energy.

## 3    Parameters Discovery

### 3.1   Overview

Table 1 lists the names of the 15 discovered parameters in Column 2 sorted by the four macro-parameters, AI Behaviour & Governance, Technology, Design & Development, and Operations. These macro-parameters will be discussed in Sections 3.2 to 3.5. The table provides one example research work for each parameter along with its research dimension (Column 3), the AI behaviour addressed by each work (Column 4) and the summary of the work (Column 5). Section 3.6 provides the taxonomy and temporal progression of the parameters.

**Table 1 AI Explainability and Governance in Smart Energy Systems: A Summary**

| Work | Parameter | Dimension | AI Behaviour | Summary |
|---|---|---|---|---|
| **AI Behaviour & Governance** | | | | |
| (Volkova *et al.*, 2022) | AI Behaviour | Power Services | Responsibility | Responsibility and accountability of AI in power services for monitoring smart grid performance. |
| **Technology** | | | | |
| (Haseeb *et al.*, 2022) | IoT & Edge | IoT Network | Trustworthiness | Improve data exchange for mobile sensors to increase the energy efficiency and trustworthiness of IoT networks |
| (Nemer *et al.*, 2022) | Unmanned Aerial Vehicles (UAVs) | Energy Consumption | Fairness | Distributed control of UAVs for enhancing the degree of coverage with limited and fair consumption of energy |
| (Yang *et al.*, 2022) | Blockchain | Energy Demand | Trustworthiness | Using trustworthy blockchain-based Federated Learning to fight malicious devices with great efficacy and low energy demand |
| (Kolangiappan and Kumar, 2022) | Sensor Networks | Energy Efficiency | Trustworthiness | Reliable and trustworthy deep learning method to detect black-hole attacks in wireless sensors and maximize energy efficiency |
| **Design & Development** | | | | |
| (Lee *et al.*, 2020) | Materials for Energy Storage & Systems | Battery Energy | Trustworthiness | Trustworthy approach using deep learning for data evaluation of battery energy storage systems |
| (Du, Pablos and Tywoniuk, 2021) | Physics of Energy Systems | Wasted Energy | Interpretability | Classify and interpret the wasted energy of high-energy jets |
| (D'amore *et al.*, 2022) | Sustainable Energy & Climate | Sustainable Development | Sustainability | The significance of AI in promoting sustainable development in water, food, and energy industries |
| **Operations** | | | | |
| (Sun *et al.*, 2022) | Energy Markets & Management | Energy Market | Fairness | Optimal multi-agent energy management for interconnected energy systems in the context of a co-trading market to promote fair commerce and to maintain the privacy of entities |
| (Xie, Ueda and Sugiyama, 2021) | Energy Demand Forecasting | Short-Term Load Prediction | Interpretability | Applying a two-stage interpretable model for short-term energy load prediction in power system management |
| (del Campo-Ávila *et al.*, 2021) | Solar Energy Systems | Solar Forecasting | Interpretability, Trustworthiness | Interpretable and trustworthy approach for global solar radiation forecasting |
| (Ardito *et al.*, 2022) | Anomaly Detection & Security | Fault Diagnosis | Interpretability | Applying interpretable AI methods to achieve transparency of fault diagnosis in electrical grids |
| (Manfren, James and Tronchin, 2022) | Energy-Efficient Buildings | Energy Consumption | Interpretability | The generalizability of interpretable machine learning for estimating building energy consumption and make buildings more energy efficient |
| (Luo *et al.*, 2021) | Grid Reliability & Stability Management | Power Stability | Interpretability | Improve the reliability of smart grids' short-term voltage stability evaluation to avoid power interruptions by using interpretable ML |
| (Chen *et al.*, 2019) | Smart City Energy Systems | Energy Meters | Interpretability | Deployment of smart energy meters for smart homes using AI-interpretable cloud analytics |

## 3.2    AI Behaviour & Governance

This parameter is about the governance and management of AI in the energy sector by identifying the requirements to build ethical, responsible, trustworthy applications and to discuss its policies, regulations, and data privacy concerns. It captures various dimensions of AI behaviour and governance including AI responsibility and accountability in smart grids (Volkova *et al.*, 2022), AI governance and regulations in power-related general-purpose technologies (Nitzberg and Zysman, 2022), reviewing the European law for the governance of AI in the electricity sector in order to allow transparent and responsible grid management (Niet, van Est and Veraart, 2021), promoting fairness and consumer protection via the use of automated decision-making to get access to fundamental services such as electricity and telecommunications (Przhedetsky, 2021), ethics of AI and power systems.

## 3.3    Technology

### 3.3.1 IoT & Edge

This parameter is about the use of the Internet of Things (IoT) and edge computing for energy systems' monitoring and efficient governance. It captures various dimensions of  "IoT & Edge," including detecting power consumption attacks for promoting vehicular edge devices reliability and AI chips' trustworthiness (Zhu *et al.*, 2022), enabling sustainable energy and ethical stable power applications by self-powered, learning sensor systems (Alagumalai *et al.*, 2022), and improving the data exchange for mobile sensors to increase the energy efficiency and trustworthiness of the IoT network (Haseeb *et al.*, 2022). Additional dimensions include efficiency in energy use via trustworthy intelligent IoT environments (Soret *et al.*, 2022), cloud computing to monitor Wireless Sensor Network (WSNs), cloud and edge computing in energy systems applications, IoT devices in the power network, edge-cloud computing in energy monitoring, and edge computing for IoT energy systems.

### 3.3.2 Unmanned Aerial Vehicles (UAVs)

This parameter involves the design of fair and trustworthy solutions to support the management and resource allocation in smart energy systems using UAVs. It captures different dimensions of "Unmanned Aerial Vehicles (UAVs)," including distributed control of UAVs for enhancing the degree of coverage with limited and fair consumption of energy (Nemer *et al.*, 2022), a fair design for multi-UAV pathway (Zhang *et al.*, 2022), UAVs trajectory design and time allocation for fair communication in wireless NOMA-IoT networks (Zhang, Xu and Wu, 2022), and fair wireless communication in UAV base stations (Qin *et al.*, 2021). Other dimensions include a fairness methodology to federated learning in vehicular edge computing (Xiao *et al.*, 2022), resource allocation in 5G Integrated Backhaul and Access (IAB) networks to increase the trustworthiness of access links (Huang C., Wang X. and Wang X., no date), and allocating resources across multiple UAVs in the IoT networks using a Deep Learning (DL) approach (Seid *et al.*, 2021).

### 3.3.3 Blockchain

This parameter is about improving the performance of AI applications for smart grids by integrating them with IoT and Blockchain technologies to obtain reliable and fair solutions. It captures various dimensions of "Blockchain," including the management of electricity demand in smart grids using blockchain as a trustworthy platform (Jose *et al.*, 2022), accountability and fairness in the energy environment using blockchain (Baashar *et al.*, 2021), and using trustworthy blockchain-based federated learning to fight malicious devices with significant efficacy and low energy demand (Yang *et al.*, 2022). Further dimensions include reliable, fair, and secure solutions for energy applications using blockchain (Al-Abri *et al.*, 2022), providing data privacy and fairness via the use of blockchain and AI-powered IoT for energy management and power trading (Lin *et al.*, 2022), a blockchain-based framework for privacy and security in energy networks, and attack detection.

### 3.3.4 Sensor Networks (SN)

This parameter is about adopting trustworthy systems to enhance the energy efficiency, lifetime, and performance of WSNs for smart grids. It captures various dimensions of "Senser Networks," including task off-loading at smart grids' edge for trustworthiness (Gunaratne *et al.*, 2022), enhancing the effectiveness of IoT-WSN by using the Secure Energy-Aware Meta-Heuristic Routing (SEAMHR) protocol (Gurram, Shariff and Biradar, 2022), and reliable and trustworthy DL method to detect black-hole attacks in wireless sensors and maximize energy efficiency (Kolangiappan and Kumar, 2022). Further dimensions include temperature-aware trustworthy routing in SNs to optimize energy efficiency (Khan *et al.*, 2022) and DL techniques to improve energy usage fairness across the cluster members in Cognitive Radio Sensor Networks (CRSN) (Stephan *et al.*, 2021).

## 3.4    Design & Development

### 3.4.1 Materials for Energy Storage & Systems

This parameter involves adopting machine learning interpretable methodologies for analyzing characteristics of chemical materials, energy systems, and batteries. It captures various dimensions of "Materials for Energy Storage & Systems," including using IML for estimating decomposition enthalpy that measures the durability of Chevrel phases for batteries (Singstock *et al.*, 2021), forecasting material characteristics using IML models to provide transparency (Allen and Tkatchenko, 2022), and trustworthy approach using DL for data evaluation of battery energy storage systems (Lee *et al.*, 2020). Additional dimensions comprise modeling and explainability of the formation energy of inorganic chemicals using DL (Huang and Ling, 2020) and developing a predictive model using IML to predict the Fermi energy level needed to build electrically conductive materials, heterostructures, and devices (Motevalli, Fox and Barnard, 2022).

### 3.4.2 Physics of Energy Systems

This parameter is about developing explainable and reliable ML models and Deep Neural Networks (DNNs) in energy systems physics. It captures various dimensions of  the "Physics of Energy System," involving developing reusable and fair intelligent systems in high-energy particle physics (Chen *et al.*, 2021), IML methodology to improve propulsion and power systems (Longmire and Banuti, 2022), classifying and interpreting the wasted energy of high-energy jets (Du, Pablos and Tywoniuk, 2021), ML approach in Kondo physics to optimize explainability (Miles *et al.*, 2021), and the reliability of semiconductors (Amrouch *et al.*, 2021).

### 3.4.3 Sustainable Energy & Climate

This parameter is about investigating AI governance's role in promoting sustainable energy and sustainable development without putting essential energy requirements at risk and considering strategies to fight climate change. It captures different dimensions of "Sustainable Energy & Climate," including AI-powered solutions to achieve sustainable energy (Saheb, Dehghani and Saheb, 2022), the aspects of water governance in urban areas (Goulas *et al.*, 2022), the significance of AI in promoting sustainable development in the water, food, and energy industries (D'amore *et al.*, 2022), sustainable education and society in the energy domain (Skowronek *et al.*, 2022), and the governance of AI to confront climate change and achieve sustainable development (Raper *et al.*, 2022).

## 3.5 Operations

### 3.5.1 Energy Markets & Management

This parameter is about detecting and governing the power demand level in energy markets. It captures various dimensions of "Energy Markets & Management," including using the IML method for the management of decentralized optimal power flow (Serna Torre and Hidalgo-Gonzalez, 2022), designing an interpretable Deep reinforcement learning (DRL) approach for transmission network expansion in wind power (Y. Wang *et al.*, 2021), and power distribution systems' reliability, interpretability, and security (Gao and Yu, 2021). Further dimensions include optimal multi-agent energy management for interconnected energy systems in the context of a co-trading market to promote fair commerce and to maintain the privacy of entities (Sun *et al.*, 2022), management of energy pipeline infrastructure (Belinsky and Afanasev, 2021), and applying IML and collaborative game theory for market regression analysis and its use in energy forecasting (Pinson, Han and Kazempour, 2021).

### 3.5.2 Energy Demand Forecasting

This parameter is about data analysis to predict energy consumption and the costs for its associated services. It captures different dimensions of "Energy Demand Forecasting," including the application of XAI in the assessment of power grid control (Kruse, Schäfer and Witthaut, 2022), employing Recurrent Neural Network (RNN) explainable method to predict short-term electric demands (Gürses-Tran, Körner and Monti, 2022), and interpretability for forecasting of probabilistic load in power network (Arora *et al.*, 2022). Moreover, using a two-stage interpretable model for short-term energy load prediction in power management (Xie, Ueda and Sugiyama, 2021), improves the effectiveness of energy resource management and increases the accuracy of forecasting power consumption over the short term with IML models (Sujan Reddy *et al.*, 2022). Additional dimensions include a multi-step interpretable probabilistic model for predicting residential energy consumption (Xu, Li and Zhou, 2022), short-term energy forecasting, energy consumption forecasting, load forecasting, demand forecasting, and various machine learning models for energy forecasting.

### 3.5.3 Solar Energy Systems

This parameter is about solar energy forecasting to enhance the management of power generation and propose trustworthy and explainable approaches. It captures various dimensions of "Solar Energy Systems," including the interpretability of solar energy forecasting (Liu *et al.*, 2022), the prediction of photovoltaic (PV) power generation that includes interpretable temporal dynamics (López Santos *et al.*, 2022), and interpretable and trustworthy approach for global solar radiation forecasting (del Campo-Ávila *et al.*, 2021), predicted energy output and $CO_2$ emissions (Bouziane and Khadir, 2022), a hybrid approach involving ML and IoT for solar radiation prediction (Ghosh *et al.*, 2020), and irradiance in solar energy, power generation by solar energy.

### 3.5.4 Anomaly Detection & Security

This parameter is about detecting, monitoring, and classifying faults and security threats in smart energy systems using transparent and knowledge-based methods. It captures various dimensions of "Anomaly Detection & Security," including applying XAI methods to identify conductive galloping in power grids (Sun *et al.*, 2021), achieve transparency of fault diagnosis in electrical grids (Ardito *et al.*, 2022), and monitoring data poisoning attacks in smart grids using white-box and black-box analysis (Bhattacharjee, Islam and Abedzadeh, 2022). Other dimensions include identifying erroneous measurements in the smart grid measuring system using trustworthy data sources (Badr *et al.*, 2022), improving DDoS security of SDN-based smart grids to increase security and trustworthiness (Nagaraj, Starke and McNair, 2021), fault detection (Landwehr *et al.*, 2022), anomaly classification for power consumption data in smart grid (Bhattacharjee, Madhavarapu and Das, 2021), anomaly attacks detection in power networks, and ML and DL models for fault detection in power systems.

### 3.5.5 Energy-Efficient Buildings

This parameter is related to adopting reliable AI models for effective management and accurate building energy consumption forecast. It captures different dimensions of "Energy-Efficient Buildings," including the generalizability of IML for estimating building energy consumption and making buildings more energy efficient (Manfren, James and Tronchin, 2022), classification of building energy performance certificates using XAI (Tsoka *et al.*, 2022), and XAI approach for forecasting long-term building energy consumption (Wenninger, Kaymakci and Wiethe, 2022). Further dimensions include providing smart recommendations based on XAI for evaluating building energy efficiency systems (Himeur *et al.*, 2022), improving the effective use of energy by adopting an IML model to forecast room occupancy (Abdel-Razek *et al.*, 2022), predicting energy consumption in buildings, and machine learning-based models for energy efficiency in buildings.

### 3.5.6 Grid Reliability & Stability Management

This parameter is about improving energy system operations via reliable assessment models and enhancing transparency in decision-making to assure supply security and system integrity. It captures dimensions of "Grid Reliability & Stability Management," including investigating the hazards of operating a state power grid to support responsible decision-making and management (Zhang *et al.*, 2023) and applying ML to improve the performance of nuclear power plants. Other dimensions include discussing the ability to apply interpretable solutions (Volodin and Tolokonskij, 2022) and designing a data-integration model to forecast the frequency response of a power system to help enhance the interpretability of the outcomes (X. Wang *et al.*, 2021). Moreover, reviewing the future of AI applications in the power system to support interpretability and stability (Zhao *et al.*, 2021), improve the reliability of smart grid's short-term voltage stability evaluation to avoid power interruptions (Luo *et al.*, 2021), power system stability, and accuracy of ML methods for the assessment of power system, and evaluating the performance of ML methods for fault selection in power lines depending on the accuracy and explainability (Gutierrez-Rojas *et al.*, 2022).

### 3.5.7 Smart City Energy Systems

This parameter involves using AI and IoT technologies to enhance the governance of smart cities systems and applications for energy. It captures various dimensions of "Smart City Energy Systems," including the usage of edge AI and blockchain to manage vehicle surveillance and traffic congestion through trustworthy automobile communications and smart energy trading (Bracco *et al.*, 2022), AI technologies in smart city governance to boost the innovative value and measurable efficiency of smart grids, electric cars, and smart buildings (Zamponi and Barbierato, 2022), and examine the most recent strategies for integrating AI and Analytics (AIA) into smart grid developments in order to enhance energy governance (Khosrojerdi *et al.*, 2022). Additional dimensions include AI-based fairness methods for transportation localization utilizing sustainable standards (Kleisarchaki *et al.*, 2022), smart city governance and planning using AI-based applications, such as smart transportation, smart education, and smart grid (Ashwini, Savithramma and Sumathi, 2022), applying DL to smart city environments and power forecasting (Naoui *et al.*, 2021), IoT applications in smart cities such as smart transportation, smart energy, and smart governance (Ilyas, 2021), and deployment of smart energy meters for smart homes using AI-interpretable cloud analytics (Chen *et al.*, 2019).

### 3.6    Taxonomy & Temporal Progression

Figure 1 depicts the taxonomy of AI governance of energy systems. Figure 2 to Figure 5 provide the temporal progression of all the 15 parameters grouped into four macro-parameters. The overall intensity of research in each macro-parameter can be seen by integrating the research of its parameters.

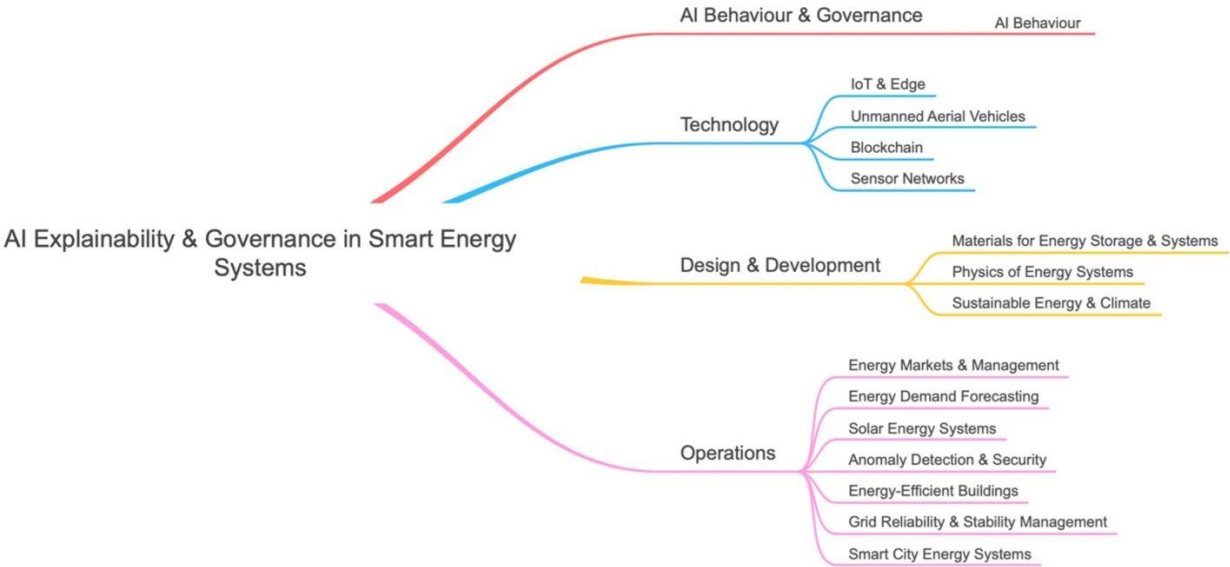

**Figure 1 Taxonomy of AI Explainability and Governance in Smart Energy Systems (1967 - 2022)**

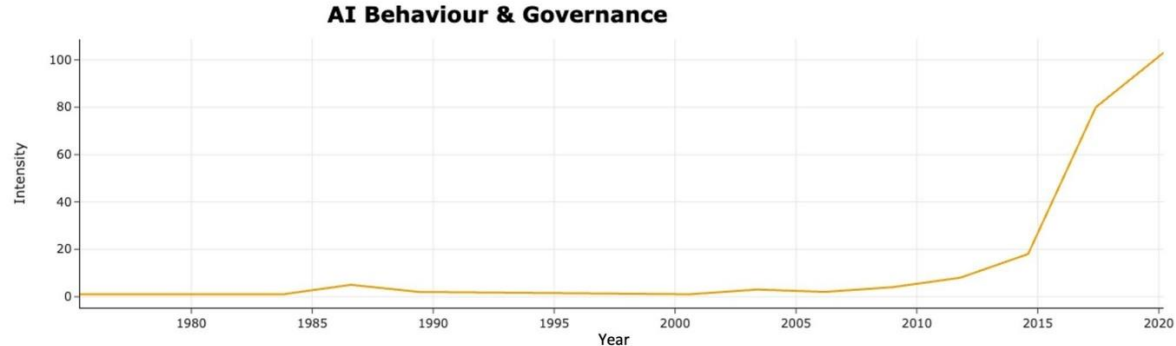

**Figure 2 Temporal Progression (AI Behaviour & Governance)**

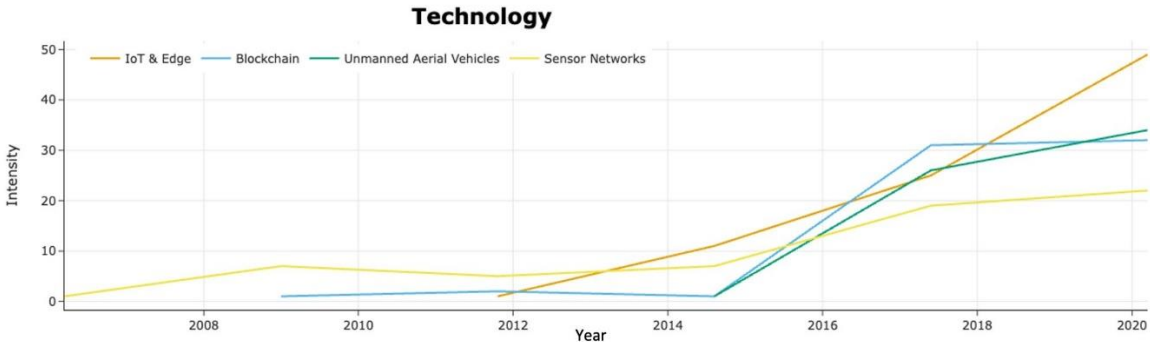

**Figure 3 Temporal Progression (Technology)**

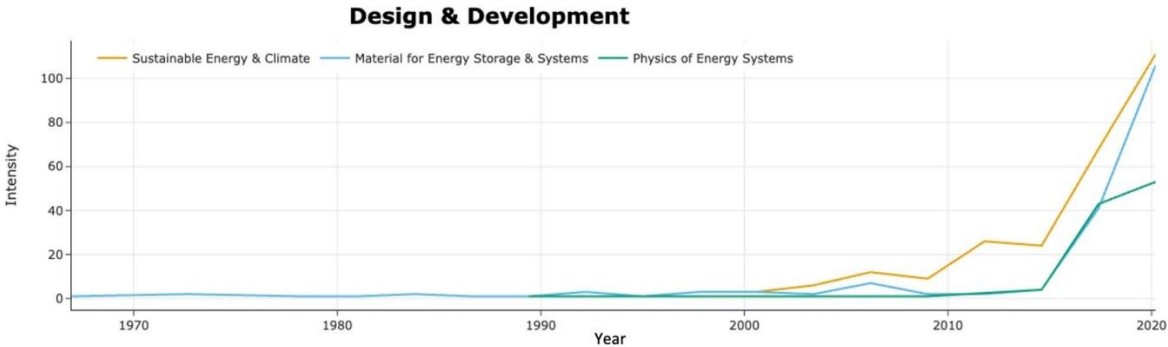

**Figure 4 Temporal Progression (Design & Development)**

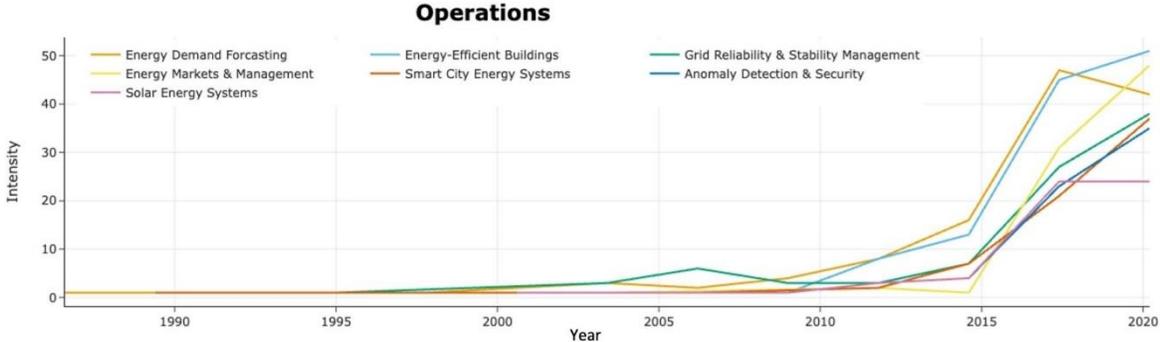

**Figure 5 Temporal Progression (Energy System Operations)**

### 3.7    Research in AI and Energy Systems (1967 - 2022)

Finally, we provide here an overview of research on AI and energy systems carried out between 1967 and 2022 without regard to the discovered parameters. Our methodology has afforded us a means for developing a comprehensive dataset and comprehension of research on AI in energy systems.

**1967 - 1989:** There are a total of 31 works in this period. These include AI, ethics, and human responsibility (Boden, 1987) and expert systems for reliable electric power (Siddiqi and Lubkeman, 1988). Some of the collected works are not related to energy or AI and fell under the search due to key terms such as power, responsibility, and reliability, for instance, responsible energy-rich biological compounds (Hager, 1967), ethical responsibility of automation (George, 1976), responsible decision-making in mental health organizations (Kochen, 1975), responsible AI in military applications (Beusmans and Wieckert, 1989), conserving power in robots (Selfridge and Franklin, 1990), and responsible utilization of AI (Stamper, 1988).

**1990 - 1999:** There are a total of 47 works in this period. These include the impact of AI on power plant reliability (Christie, 1990), power management and distribution for automation (Ashworth, 1990), nuclear power plant (Trovato, Imai and Touchton, 1990), explainability of electric circuits (Kashihara *et al.*, 1992), AI in power systems management (Janik and Gholdston, 1992; Hasan, Ramsay and Moyes, 1994; Irisarri, 1996), short-term electric load forecasting (Stroemich and Thomas, 1997), reliability of nuclear weapons stockpile protection measures (Molley, 1996), AI in fault diagnosis systems for power plants (Kiupel *et al.*, 1995), electricity transportation management (Jennings, 1995), decision tree interpretability to manage electric power utilities (Wehenkel *et al.*, 1994), explainability for the evaluation of power system security (Boyen and Wehenkel, 1999), and trustworthy intelligent agents to power markets (Krishna and Ramesh, 1998).

**2000 - 2009:** There are a total of 168 works in this period. These include electrical stimulation systems (Fisekovic and Popovic, 2001), ethical implications and responsibility of AI (Perri, 2001), improving quality of electricity (*The engagements of the management of the French transmission system (RTE, R seau de Transport Fran ais] in the matter of quality in providing electricity*, no date), fair allocation of neural networks (Fidalgo, Torres and Matos, 2007), automatic diagnosis system for power transformers using artificial neural networks (Moreira, Santos and Vellasco, 2007), engineering ethics (Berne, 2001), and renewable energy applications (Belu, 2009).

**2010 - 2019:** There are a total of 1142 works in this period. These include energy consumption in wireless sensor networks (Ebadi *et al.*, 2010), reliability and efficiency of smart grids (Rosic, Novak and Vukmirovic, 2013), simulation of electric power markets (Vale *et al.*, 2011), social responsibility for low energy consumption and public building energy management (Egging, 2013), smart grid IT governance (Parra, Arroyo and Garcia, 2014), energy marketsagent-based modelling (Lupo and Kiprakis, 2015), cybersecurity for smart grids (Yardley *et al.*, 2015), solar power prediction (Cabrera, Benhaddou and Ordonez, 2016), pricing systems in smart grids (Bandyopadhyay *et al.*, 2016), smart grid congestion management (MacDougall *et al.*, 2017), electricity theft in smart grids (Yeckle and Tang, 2018), and load scheduling in smart grids (Senevirathne *et al.*, 2019).

**2020 - 2022:** There are a total of 2180 works during this less than a three-year period. We can say that the last three years have seen a surge in AI explainability research in energy systems. The works in this period are based on the latest trends in ML and DL, explainability and interpretability, and smart approaches toward multi-source energy systems.

## 4 Discussion and Conclusion

This paper provided a review of AI explainability and governance in smart energy systems. We automatically discovered 15 parameters and elaborated the research landscape by reviewing over 150 papers. The parameters were grouped into four macro-parameters, namely AI Behaviour & Governance, Technology, Design & Development, and Operations.

Our work supports and extends the existing literature, particularly (Machlev *et al.*, 2022; Xu *et al.*, 2022), that identified stability, reliability, energy forecasting, and power system flexibility as major activities in the field. This work has provided an extensive view of AI governance in energy systems and thereby has broadened and deepened the understanding of the field.

The work has identified a range of specific and broad challenges including resource allocation in wireless sensor networks with multiple UAVs (Seid *et al.*, 2021; Zhang, Xu and Wu, 2022), governance of AI in power-related general-purpose technologies (Niet, van Est and Veraart, 2021; Przhedetsky, 2021; Nitzberg and Zysman, 2022), fault detection, fault diagnosis, and anomaly detection in smart energy systems (Sun *et al.*, 2021; Badr *et al.*, 2022), edge computing for detecting power demand attacks (Alagumalai *et al.*, 2022; Haseeb *et al.*, 2022; Zhu *et al.*, 2022), blockchain-based reliability and security (Al-Abri *et al.*, 2022; Jose *et al.*, 2022), governance of energy markets and energy pipeline systems (Belinsky and Afanasev, 2021; Serna Torre and Hidalgo-Gonzalez, 2022; Sun *et al.*, 2022), forecasting short-term energy demand (Xie, Ueda and Sugiyama, 2021; Gürses-Tran, Körner and Monti, 2022), energy trading using federated learning in smart cities (Bracco *et al.*, 2022), energy-saving edge AI applications (Khosrojerdi *et al.*, 2022), performance optimization and stability of smart grid operations and nuclear power systems (Luo *et al.*, 2021; Volodin and Tolokonskij, 2022), and others. All these areas are candidates for future research.

We have listed above a range of challenges related to explainable AI in energy systems. It is apparent that both explainable AI and energy systems are rich and complex fields. A proper discussion of specific limitations and challenges on the subject requires several pages. Due to the lack of space, we briefly mention below a few examples of specific challenges in explainable AI for energy systems. Fault detection, diagnosis, and prediction are among the most important challenges in energy systems. These are problematic due to the complexity of energy systems covering large and uninhabited geographic regions involving difficult terrains. Specific fault detection-related challenges include the management and storage issues arising due to the use of multiple data sources (solar or wind power forecasting and related faults using numerical and image data), as opposed to a single data source, for fault detection (Landwehr *et al.*, 2022), the influence of measurement noise on fault prediction performance (Sun *et al.*, 2021), privacy issues in fault-diagnosis and examination of security and stability-sensitive scenarios (Ardito *et al.*, 2022), and low accuracies of AI algorithms in fault detection, diagnosis, and prediction (Wu *et al.*, 2022). Another increasingly important area is the security of ML and DL software (Altoub *et al.*, 2022). The challenges in this area include, among others, data poisoning attacks and the performance of related solutions (Bhattacharjee, Islam and Abedzadeh, 2022), and anomaly detection methods for smart grid meter security against poisoning attacks (Bhattacharjee, Madhavarapu and Das, 2021).

The parameters discovery shows that most of the research is focussed on Operations followed by research activities in Design and Technology. The least research is in AI Behaviour & Governance where much effort is needed in the future. The methods and tools to support trustworthiness (explainability and other AI traits) in AI for energy systems include, among others, visual explanation techniques using Grad-CAM (Ardito *et al.*, 2022), sequence-to-sequence RNN methods for visual explanation of short-term load forecasting (Gürses-Tran, Körner and Monti, 2022), the Scale-Invariant Feature Transform (SIFT) method (Singstock *et al.*, 2021), post-hoc interpretability (Allen and Tkatchenko, 2022), SHapley Additive exPlanation (SHAP) (Pinson, Han and Kazempour, 2021; Abdel-Razek *et al.*, 2022; Kruse, Schäfer and Witthaut, 2022), interpretable Tiny Neural Networks (TNN) (Longmire and Banuti, 2022), model-agnostic methods (Gürses-Tran, Körner and Monti, 2022), the use of Temporal Fusion Transformer (TFT) method to enhance interpretability (López Santos *et al.*, 2022), the decision tree and Classification and Regression Tree (CART) algorithms for ML explainability (Sun *et al.*, 2021), visual data exploration for the interpretability of fault diagnosis (Landwehr *et al.*, 2022), a partially interpretable method using LSTM and MLP (multilayer perceptron) for short-term load forecasting (Xie, Ueda and Sugiyama, 2021), and Local Interpretable Model-Agnostic Explanation (LIME) (Tsoka *et al.*, 2022). We expect that many more methods will be developed for explainable AI in the future.

Note that the review and analysis presented in this paper are based on the works indexed in the Scopus database. Incorporating other databases in our deep journalism tool is expected to create additional parameters and structure of research on AI in energy systems. Future work will investigate the use of our deep journalism tool with additional research databases.

## 5    Conflict of Interest

*The authors declare that the research was conducted in the absence of any commercial or financial relationships that could be construed as a potential conflict of interest.*

## 6    Author Contributions

RAS and RM conceived, developed, analysed, and validated the study. RAS developed the software. RAS and RM prepared the initial draft, reviewed and edited by RM and IK. RM and IK provided supervision, funds, resources, and contributed to the article editing.

## 7    Funding

The authors acknowledge with thanks the technical and financial support from the Deanship of Scientific Research (DSR) at the King Abdulaziz University (KAU), Jeddah, Saudi Arabia, under Grant No. RG-11-611-38.

## 8    Acknowledgments

The work carried out in this paper is supported by the HPC Center at the King Abdulaziz University.

## 9    References

Abdel-Razek, S. A. *et al.* (2022) 'Energy Efficiency through the Implementation of an AI Model to Predict Room Occupancy Based on Thermal Comfort Parameters', *Sustainability 2022, Vol. 14, Page 7734*, 14(13), p. 7734. doi: 10.3390/SU14137734.

Ahmad, I. *et al.* (2022) 'Deep Journalism and DeepJournal V1.0: A Data-Driven Deep Learning

Approach to Discover Parameters for Transportation', *Sustainability (Switzerland)*, 14(9), p. 5711. doi: 10.3390/SU14095711.

Al-Abri, T. *et al.* (2022) 'Review on Energy Application Using Blockchain Technology with an Introductions in the Pricing Infrastructure', *IEEE Access*, 10, pp. 80119–80137. doi: 10.1109/ACCESS.2022.3194161.

Alagumalai, A. *et al.* (2022) 'Self-powered sensing systems with learning capability', *Joule*, 6(7), pp. 1475–1500. doi: 10.1016/J.JOULE.2022.06.001.

Alahmari, N. *et al.* (2022) 'Musawah: A Data-Driven AI Approach and Tool to Co-Create Healthcare Services with a Case Study on Cancer Disease in Saudi Arabia', *Sustainability 2022, Vol. 14, Page 3313*, 14(6), p. 3313. doi: 10.3390/SU14063313.

Alkhayat, G., Hasan, S. H. and Mehmood, R. (2022) 'SENERGY: A Novel Deep Learning-Based Auto-Selective Approach and Tool for Solar Energy Forecasting', *Energies 2022, Vol. 15, Page 6659*, 15(18), p. 6659. doi: 10.3390/EN15186659.

Allen, A. E. A. and Tkatchenko, A. (2022) 'Machine learning of material properties: Predictive and interpretable multilinear models', *Science Advances*, 8(18), p. 7185. doi: 10.1126/SCIADV.ABM7185/SUPPL_FILE/SCIADV.ABM7185_SM.PDF.

Alomari, E. *et al.* (2021) 'Iktishaf+: A Big Data Tool with Automatic Labeling for Road Traffic Social Sensing and Event Detection Using Distributed Machine Learning', *Sensors*, 21(9), p. 2993. doi: 10.3390/s21092993.

Alqahtani, E. *et al.* (2022) 'Smart Homes and Families to Enable Sustainable Societies: A Data-Driven Approach for Multi-Perspective Parameter Discovery using BERT Modelling', *Preprints*. doi: 10.20944/PREPRINTS202208.0233.V1.

Alswedani, S. *et al.* (2022) 'Discovering Urban Governance Parameters for Online Learning in Saudi Arabia During COVID-19 Using Topic Modeling of Twitter Data', *Frontiers in Sustainable Cities*, 4, pp. 1–24. doi: 10.3389/FRSC.2022.751681.

Altoub, M. *et al.* (2022) 'An Ontological Knowledge Base of Poisoning Attacks on Deep Neural Networks', *Applied Sciences 2022, Vol. 12, Page 11053*, 12(21), p. 11053. doi: 10.3390/APP122111053.

Amrouch, H. *et al.* (2021) 'Special session: Machine learning for semiconductor test and reliability', *Proceedings of the IEEE VLSI Test Symposium*, 2021-April. doi: 10.1109/VTS50974.2021.9441052.

Ardito, C. *et al.* (2022) 'Visual inspection of fault type and zone prediction in electrical grids using interpretable spectrogram-based CNN modeling', *Expert Systems with Applications*, 210, p. 118368. doi: 10.1016/J.ESWA.2022.118368.

Arora, P. *et al.* (2022) 'Remodelling State-Space Prediction with Deep Neural Networks for Probabilistic Load Forecasting', *IEEE Transactions on Emerging Topics in Computational Intelligence*, 6(3), pp. 628–637. doi: 10.1109/TETCI.2021.3064028.

Ashwini, B. P., Savithramma, R. M. and Sumathi, R. (2022) 'Artificial Intelligence in Smart City

Applications: An overview', *Proceedings - 2022 6th International Conference on Intelligent Computing and Control Systems, ICICCS 2022*, pp. 986–993. doi: 10.1109/ICICCS53718.2022.9788152.

Ashworth, B. R. (1990) 'Managing autonomy levels in the SSM/PMAD testbed', *Proceedings of the Intersociety Energy Conversion Engineering Conference*, 1, pp. 263–268. doi: 10.1109/IECEC.1990.716891.

Baashar, Y. *et al.* (2021) 'Toward Blockchain Technology in the Energy Environment', *Sustainability 2021, Vol. 13, Page 9008*, 13(16), p. 9008. doi: 10.3390/SU13169008.

Badr, M. M. *et al.* (2022) 'Detection of False-Reading Attacks in Smart Grid Net-Metering System', *IEEE Internet of Things Journal*, 9(2), pp. 1386–1401. doi: 10.1109/JIOT.2021.3087580.

Bandyopadhyay, S. *et al.* (2016) 'An axiomatic framework for ex-ante dynamic pricing mechanisms in smart grid', *30th AAAI Conference on Artificial Intelligence, AAAI 2016*, pp. 3800–3806. doi: 10.1609/aaai.v30i1.9900.

Belinsky, A. and Afanasev, V. (2021) 'Optimal Control of Energy Pipeline Systems Based on Deep Reinforcement Learning', *Lecture Notes in Networks and Systems*, 155, pp. 1348–1355. doi: 10.1007/978-3-030-59126-7_148.

Belu, R. (2009) 'A project-based power electronics course with an increased content of renewable energy applications', *ASEE Annual Conference and Exposition, Conference Proceedings*. doi: 10.18260/1-2--4994.

Berne, R. W. (2001) 'Reaching and teaching through "The Matrix"; robosapiens, transhumanism, and the formidable in engineering ethics', *ASEE Annual Conference Proceedings*, pp. 8343–8350. doi: 10.18260/1-2--9710.

Beusmans, J. and Wieckert, K. (1989) 'Computing, research, and war: if knowledge is power, where is responsibility?', *Communications of the ACM*, 32(8), pp. 939–951. doi: 10.1145/65971.65973.

Bhattacharjee, S., Islam, M. J. and Abedzadeh, S. (2022) 'Robust Anomaly based Attack Detection in Smart Grids under Data Poisoning Attacks', *CPSS 2022 - Proceedings of the 8th ACM Cyber-Physical System Security Workshop*, pp. 3–14. doi: 10.1145/3494107.3522778.

Bhattacharjee, S., Madhavarapu, P. and Das, S. K. (2021) 'A Diversity Index based Scoring Framework for Identifying Smart Meters Launching Stealthy Data Falsification Attacks', *ASIA CCS 2021 - Proceedings of the 2021 ACM Asia Conference on Computer and Communications Security*, pp. 26–39. doi: 10.1145/3433210.3437527.

Blondeel, M. *et al.* (2021) 'The geopolitics of energy system transformation: A review', *Geography Compass*, 15(7), p. e12580. doi: 10.1111/GEC3.12580.

Boden, M. (1987) 'Artificial intelligence: Cannibal or missionary?', *AI & SOCIETY 1987 1:1*, 1(1), pp. 17–23. doi: 10.1007/BF01905886.

Bouziane, S. E. and Khadir, M. T. (2022) 'Towards an energy management system based on a multi-agent architecture and LSTM networks', *https://doi.org/10.1080/0952813X.2022.2093407*. doi:

10.1080/0952813X.2022.2093407.

Boyen, X. and Wehenkel, L. (1999) 'Automatic induction of fuzzy decision trees and its application to power system security assessment', *Fuzzy Sets and Systems*, 102(1), pp. 3–19. doi: 10.1016/S0165-0114(98)00198-5.

Bracco, S. *et al.* (2022) 'Edge AI and Blockchain for Smart Sustainable Cities: Promise and Potential', *Sustainability 2022, Vol. 14, Page 7609*, 14(13), p. 7609. doi: 10.3390/SU14137609.

Cabrera, W., Benhaddou, D. and Ordonez, C. (2016) 'Solar Power Prediction for Smart Community Microgrid', *2016 IEEE International Conference on Smart Computing, SMARTCOMP 2016*. doi: 10.1109/SMARTCOMP.2016.7501718.

del Campo-Ávila, J. *et al.* (2021) 'Binding data mining and expert knowledge for one-day-ahead prediction of hourly global solar radiation', *Expert Systems with Applications*, 167, p. 114147. doi: 10.1016/J.ESWA.2020.114147.

Castelvecchi, D. (2016) 'Can we open the black box of AI?', *Nature International Weekly Journal of Science*, 538(7623), pp. 20–23. doi: 10.1038/538020a.

Ceglia, F. *et al.* (2020) 'From smart energy community to smart energy municipalities: Literature review, agendas and pathways', *Journal of Cleaner Production*, 254, p. 120118. doi: 10.1016/J.JCLEPRO.2020.120118.

Chen, Y. *et al.* (2021) 'A FAIR and AI-ready Higgs boson decay dataset', *Scientific Data*, 9(1). doi: 10.1038/s41597-021-01109-0.

Chen, Y. Y. *et al.* (2019) 'Design and Implementation of Cloud Analytics-Assisted Smart Power Meters Considering Advanced Artificial Intelligence as Edge Analytics in Demand-Side Management for Smart Homes', *Sensors (Basel, Switzerland)*, 19(9). doi: 10.3390/S19092047.

Christie, R. D. (1990) 'Impact of artificial intelligence on plant and system operations', *Instrumentation, Control, and Automation in the Power Industry, Proceedings*, 33, pp. 193–197.

D'amore, G. *et al.* (2022) 'Artificial Intelligence in the Water–Energy–Food Model: A Holistic Approach towards Sustainable Development Goals', *Sustainability 2022, Vol. 14, Page 867*, 14(2), p. 867. doi: 10.3390/SU14020867.

Doran, D., Schulz, S. and Besold, T. R. (2017) 'What Does Explainable AI Really Mean? A New Conceptualization of Perspectives', *CEUR Workshop Proceedings*, 2071. doi: 10.48550/arxiv.1710.00794.

Du, Y.-L., Pablos, D. and Tywoniuk, K. (2021) 'Classification of quark and gluon jets in hot QCD medium with deep learning', *Proceedings of Science*, 380. doi: 10.22323/1.380.0224.

Ebadi, S. *et al.* (2010) 'Energy balancing in wireless sensor networks with selecting two cluster-heads in hierarchical clustering', *Proceedings - 2010 International Conference on Computational Intelligence and Communication Networks, CICN 2010*, pp. 230–233. doi: 10.1109/CICN.2010.55.

Egging, R. (2013) 'Drivers, trends, and uncertainty in long-term price projections for energy management in public buildings', *Energy Policy*, 62, pp. 617–624. doi: 10.1016/J.ENPOL.2013.07.022.

Fidalgo, J. N., Torres, J. A. F. M. and Matos, M. (2007) 'Fair allocation of distribution losses based on neural networks', *2007 International Conference on Intelligent Systems Applications to Power Systems, ISAP*. doi: 10.1109/ISAP.2007.4441685.

Fisekovic, N. and Popovic, D. B. (2001) 'New controller for functional electrical stimulation systems', *Medical Engineering and Physics*, 23(6), pp. 391–399. doi: 10.1016/S1350-4533(01)00069-8.

Gao, Y. and Yu, N. (2021) 'Deep reinforcement learning in power distribution systems: Overview, challenges, and opportunities', *2021 IEEE Power and Energy Society Innovative Smart Grid Technologies Conference, ISGT 2021*. doi: 10.1109/ISGT49243.2021.9372283.

George, M. (1976) 'PRESS BUTTON AGE.', *Electronics and Power*, 22(5), pp. 307–309. doi: 10.1049/EP.1976.0145.

Ghosh, S. *et al.* (2020) 'Smart Cropping based on Predicted Solar Radiation using IoT and Machine Learning', *Proceedings of IEEE International Conference on Advent Trends in Multidisciplinary Research and Innovation, ICATMRI 2020*. doi: 10.1109/ICATMRI51801.2020.9398323.

Gill-Wiehl, A. *et al.* (2022) 'Beyond customer acquisition: A comprehensive review of community participation in mini grid projects', *Renewable and Sustainable Energy Reviews*, 153, p. 111778. doi: 10.1016/J.RSER.2021.111778.

Goebel, R. *et al.* (2018) 'Explainable AI: The new 42?', *Lecture Notes in Computer Science (including subseries Lecture Notes in Artificial Intelligence and Lecture Notes in Bioinformatics)*, 11015 LNCS, pp. 295–303. doi: 10.1007/978-3-319-99740-7_21/FIGURES/6.

Goulas, A. *et al.* (2022) 'Public Perceptions of Household IoT Smart Water "Event" Meters in the UK—Implications for Urban Water Governance', *Frontiers in Sustainable Cities*, 4, p. 10. doi: 10.3389/FRSC.2022.758078/BIBTEX.

Grootendorst, M. (2021) *Interactive Topic Modeling with BERTopic | Towards Data Science*.

Gunaratne, N. G. T. *et al.* (2022) 'An Edge Tier Task Offloading to Identify Sources of Variance Shifts in Smart Grid Using a Hybrid of Wrapper and Filter Approaches', *IEEE Transactions on Green Communications and Networking*, 6(1), pp. 329–340. doi: 10.1109/TGCN.2021.3137330.

Gurram, G. V., Shariff, N. C. and Biradar, R. L. (2022) 'A Secure Energy Aware Meta-Heuristic Routing Protocol (SEAMHR) for sustainable IoT-Wireless Sensor Network (WSN)', *Theoretical Computer Science*, 930, pp. 63–76. doi: 10.1016/J.TCS.2022.07.011.

Gürses-Tran, G., Körner, T. A. and Monti, A. (2022) 'Introducing explainability in sequence-to-sequence learning for short-term load forecasting', *Electric Power Systems Research*, 212, p. 108366. doi: 10.1016/J.EPSR.2022.108366.

Gutierrez-Rojas, D. *et al.* (2022) 'Performance evaluation of machine learning for fault selection in power transmission lines', *Knowledge and Information Systems*, 64(3), pp. 859–883. doi:

10.1007/S10115-022-01657-W/FIGURES/11.

Hager, A. (1967) 'Untersuchungen über die lichtinduzierten reversiblen xanthophyllumwandlungen an Chlorella und Spinacia', *Planta 1967 74:2*, 74(2), pp. 148–172. doi: 10.1007/BF00388326.

Hagras, H. (2018) 'Toward Human-Understandable, Explainable AI', *Computer*, 51(9), pp. 28–36. doi: 10.1109/MC.2018.3620965.

Hasan, K., Ramsay, B. and Moyes, I. (1994) 'Object oriented expert systems for real-time power system alarm processing: Part I. Selection of a toolkit', *Electric Power Systems Research*, 30(1), pp. 69–75. doi: 10.1016/0378-7796(94)90061-2.

Haseeb, K. *et al.* (2022) 'Device-to-Device (D2D) Multi-Criteria Learning Algorithm Using Secured Sensors', *Sensors 2022, Vol. 22, Page 2115*, 22(6), p. 2115. doi: 10.3390/S22062115.

Himeur, Y. *et al.* (2022) 'Techno-economic assessment of building energy efficiency systems using behavioral change: A case study of an edge-based micro-moments solution', *Journal of Cleaner Production*, 331, p. 129786. doi: 10.1016/J.JCLEPRO.2021.129786.

Hoffman, R. R. *et al.* (2018) 'Metrics for Explainable AI: Challenges and Prospects'. doi: 10.48550/arxiv.1812.04608.

Huang C., Wang X. and Wang X. (no date) *Effective-Capacity-Based Resource Allocation for End-to-End Multi-Connectivity in 5G IAB Networks | IEEE Journals & Magazine | IEEE Xplore*.

Huang, L. and Ling, C. (2020) 'Practicing deep learning in materials science: An evaluation for predicting the formation energies', *Journal of Applied Physics*, 128(12), p. 124901. doi: 10.1063/5.0012411.

Hussain, A., Bui, V. H. and Kim, H. M. (2019) 'Microgrids as a resilience resource and strategies used by microgrids for enhancing resilience', *Applied Energy*, 240, pp. 56–72. doi: 10.1016/J.APENERGY.2019.02.055.

Ilyas, M. (2021) 'IoT Applications in Smart Cities', *2021 IEEE International Conference on Electronic Communications, Internet of Things and Big Data, ICEIB 2021*, pp. 44–47. doi: 10.1109/ICEIB53692.2021.9686400.

Irisarri, G. (1996) 'Integration of artificial intelligence applications in the ems: issues and solutions', *IEEE Transactions on Power Systems*, 11(1), pp. 475–482. doi: 10.1109/59.486136.

Janbi, N. *et al.* (2022) 'Imtidad: A Reference Architecture and a Case Study on Developing Distributed AI Services for Skin Disease Diagnosis over Cloud, Fog and Edge', *Sensors 2022, Vol. 22, Page 1854*, 22(5), p. 1854. doi: 10.3390/S22051854.

Janik, D. F. and Gholdston, E. W. (1992) 'A Prototype Ground Support System Security Monitor for Space Based Power System Health Monitoring', *SAE Technical Papers*. doi: 10.4271/929332.

Jennings, N. R. (1995) 'Controlling cooperative problem solving in industrial multi-agent systems using joint intentions', *Artificial Intelligence*, 75(2), pp. 195–240. doi: 10.1016/0004-3702(94)00020-2.

Jose, D. T. *et al.* (2022) 'Integrating big data and blockchain to manage energy smart grids—TOTEM framework', *Blockchain: Research and Applications*, 3(3), p. 100081. doi: 10.1016/J.BCRA.2022.100081.

Kashihara, A. *et al.* (1992) 'Advanced explanation capabilities for intelligent tutoring systems: The explanation structure model (EXSEL)', *Systems and Computers in Japan*, 23(12), pp. 93–107. doi: 10.1002/SCJ.4690231209.

Khan, T. *et al.* (2022) 'A Temperature-Aware Trusted Routing Scheme for Sensor Networks: Security Approach', *Computers & Electrical Engineering*, 98, p. 107735. doi: 10.1016/J.COMPELECENG.2022.107735.

Khosrojerdi, F. *et al.* (2022) 'Integrating artificial intelligence and analytics in smart grids: a systematic literature review', *International Journal of Energy Sector Management*, 16(2), pp. 318–338. doi: 10.1108/IJESM-06-2020-0011/FULL/XML.

Kiupel, N. *et al.* (1995) 'Fuzzy residual evaluation concept (FREC)', *Proceedings of the IEEE International Conference on Systems, Man and Cybernetics*, 1, pp. 13–18. doi: 10.1109/ICSMC.1995.537725.

Kleisarchaki, S. *et al.* (2022) 'Optimization of Soft Mobility Localization with Sustainable Policies and Open Data', *2022 18th International Conference on Intelligent Environments, IE 2022 - Proceedings*. doi: 10.1109/IE54923.2022.9826779.

Kochen, M. (1975) 'Hypothesis processing as a new tool to aid managers of mental health agencies in serving long-term regional interests', *International Journal of Bio-Medical Computing*, 6(4), pp. 299–312. doi: 10.1016/0020-7101(75)90013-6.

Kolangiappan, J. and Kumar, A. S. (2022) 'A novel framework for the prevention of black-hole in wireless sensors using hybrid convolution network', *Scientific and Technical Journal of Information Technologies, Mechanics and Optics*, 22(2), pp. 317–323. doi: 10.17586/2226-1494-2022-22-2-317-323.

Kong, P. Y. and Song, Y. (2020) 'Joint Consideration of Communication Network and Power Grid Topology for Communications in Community Smart Grid', *IEEE Transactions on Industrial Informatics*, 16(5), pp. 2895–2905. doi: 10.1109/TII.2019.2912670.

Krishna, V. and Ramesh, V. C. (1998) 'Intelligent agents for negotiations in market games, part 1: model', *IEEE Transactions on Power Systems*, 13(3), pp. 1103–1108. doi: 10.1109/59.709106.

Kruse, J., Schäfer, B. and Witthaut, D. (2022) 'Secondary control activation analysed and predicted with explainable AI', *Electric Power Systems Research*, 212, p. 108489. doi: 10.1016/J.EPSR.2022.108489.

Landwehr, J. P. *et al.* (2022) 'Design Knowledge for Deep-Learning-Enabled Image-Based Decision Support Systems: Evidence From Power Line Maintenance Decision-Making', *Business and Information Systems Engineering*, pp. 1–22. doi: 10.1007/S12599-022-00745-Z/FIGURES/9.

Lee, H. *et al.* (2020) 'Deep Learning-Based False Sensor Data Detection for Battery Energy Storage Systems', *2020 IEEE CyberPELS, CyberPELS 2020*. doi:

10.1109/CYBERPELS49534.2020.9311542.

Lin, Y. J. *et al.* (2022) 'Blockchain Power Trading and Energy Management Platform', *IEEE Access*, 10, pp. 75932–75948. doi: 10.1109/ACCESS.2022.3189472.

Liu, W. *et al.* (2022) 'Use of physics to improve solar forecast: Part II, machine learning and model interpretability', *Solar Energy*, 244, pp. 362–378. doi: 10.1016/J.SOLENER.2022.08.040.

Longmire, N. and Banuti, D. T. (2022) 'Onset of heat transfer deterioration caused by pseudo-boiling in CO2 laminar boundary layers', *International Journal of Heat and Mass Transfer*, 193, p. 122957. doi: 10.1016/J.IJHEATMASSTRANSFER.2022.122957.

López Santos, M. *et al.* (2022) 'Application of Temporal Fusion Transformer for Day-Ahead PV Power Forecasting', *Energies*, 15(14). doi: 10.3390/EN15145232.

Lundberg, S. M. *et al.* (2020) 'From local explanations to global understanding with explainable AI for trees', *Nature Machine Intelligence 2020 2:1*, 2(1), pp. 56–67. doi: 10.1038/s42256-019-0138-9.

Luo, Y. *et al.* (2021) 'Graph Convolutional Network-Based Interpretable Machine Learning Scheme in Smart Grids', *IEEE Transactions on Automation Science and Engineering*. doi: 10.1109/TASE.2021.3090671.

Lupo, S. and Kiprakis, A. (2015) 'Agent-based models for electricity markets accounting for smart grid participation', *Lecture Notes of the Institute for Computer Sciences, Social-Informatics and Telecommunications Engineering, LNICST*, 154, pp. 48–57. doi: 10.1007/978-3-319-25479-1_4.

MacDougall, P. *et al.* (2017) 'Multi-goal optimization of competing aggregators using a web-of-cells approach', *2017 IEEE PES Innovative Smart Grid Technologies Conference Europe, ISGT-Europe 2017 - Proceedings*, 2018-January, pp. 1–6. doi: 10.1109/ISGTEUROPE.2017.8260335.

Machlev, R. *et al.* (2022) 'Explainable Artificial Intelligence (XAI) techniques for energy and power systems: Review, challenges and opportunities', *Energy and AI*, 9, p. 100169. doi: 10.1016/J.EGYAI.2022.100169.

Manfren, M., James, P. A. and Tronchin, L. (2022) 'Data-driven building energy modelling – An analysis of the potential for generalisation through interpretable machine learning', *Renewable and Sustainable Energy Reviews*, 167. doi: 10.1016/J.RSER.2022.112686.

Mehmood, R. *et al.* (2017) 'UTiLearn: A Personalised Ubiquitous Teaching and Learning System for Smart Societies', *IEEE Access*, 5, pp. 2615–2635. doi: 10.1109/ACCESS.2017.2668840.

Mehmood, R. (2022) '"Deep journalism" driven by AI can aid better government', *The Mandarin*. Available at: https://www.themandarin.com.au/201467-deep-journalism-driven-by-ai-can-aid-better-government/ (Accessed: 15 October 2022).

Miles, C. *et al.* (2021) 'Machine learning of Kondo physics using variational autoencoders and symbolic regression', *Physical Review B*, 104(23). doi: 10.1103/PhysRevB.104.235111.

Molley, P. A. (1996) 'Computer vision challenges and technologies for agile manufacturing', *https://doi.org/10.1117/12.233237*, 2727, pp. 1036–1037. doi: 10.1117/12.233237.

Moreira, M. P., Santos, L. T. B. and Vellasco, M. M. B. R. (2007) 'Power transformers diagnosis using neural networks', *IEEE International Conference on Neural Networks - Conference Proceedings*, pp. 1929–1934. doi: 10.1109/IJCNN.2007.4371253.

Motevalli, B., Fox, B. L. and Barnard, A. S. (2022) 'Charge-dependent Fermi level of graphene oxide nanoflakes from machine learning', *Computational Materials Science*, 211. doi: 10.1016/J.COMMATSCI.2022.111526.

Nagaraj, K., Starke, A. and McNair, J. (2021) 'GLASS: A Graph Learning Approach for Software Defined Network Based Smart Grid DDoS Security', *IEEE International Conference on Communications*. doi: 10.1109/ICC42927.2021.9500999.

Naoui, M. A. *et al.* (2021) 'Using a distributed deep learning algorithm for analyzing big data in smart cities', *Smart and Sustainable Built Environment*, 10(1), pp. 90–105. doi: 10.1108/SASBE-04-2019-0040/FULL/XML.

Nemer, I. A. *et al.* (2022) 'Energy-Efficient UAV Movement Control for Fair Communication Coverage: A Deep Reinforcement Learning Approach', *Sensors 2022, Vol. 22, Page 1919*, 22(5), p. 1919. doi: 10.3390/S22051919.

Niet, I., van Est, R. and Veraart, F. (2021) 'Governing AI in Electricity Systems: Reflections on the EU Artificial Intelligence Bill', *Frontiers in Artificial Intelligence*, 4, p. 109. doi: 10.3389/FRAI.2021.690237/BIBTEX.

Nitzberg, M. and Zysman, J. (2022) 'Algorithms, data, and platforms: the diverse challenges of governing AI', *https://doi.org/10.1080/13501763.2022.2096668*. doi: 10.1080/13501763.2022.2096668.

Parra, I., Arroyo, G. and Garcia, A. (2014) 'Experiences and practices in the implementation of IT Governance in Mexican electric utility', *CIGRE Session 45 - 45th International Conference on Large High Voltage Electric Systems 2014*, 2014-August.

Perri (2001) 'Ethics, regulation and the new artificial intelligence, part I: Accountability and Power', *Information Communication and Society*, 4(2), pp. 199–229. doi: 10.1080/13691180110044461.

Phillips, P. J. *et al.* (2021) 'Four Principles of Explainable Artificial Intelligence', *NIST Interagency/Internal Report (NISTIR) - 8312, National Institute of Standards and Technology, Gaithersburg, MD*. doi: 10.6028/NIST.IR.8312.

Pinson, P., Han, L. and Kazempour, J. (2021) 'Regression markets and application to energy forecasting', *TOP*. doi: 10.1007/s11750-022-00631-7.

Przhedetsky, L. (2021) 'Designing Effective and Accessible Consumer Protections against Unfair Treatment in Markets where Automated Decision Making is used to Determine Access to Essential Services: A Case Study in Australia's Housing Market', *AIES 2021 - Proceedings of the 2021 AAAI/ACM Conference on AI, Ethics, and Society*, pp. 279–280. doi: 10.1145/3461702.3462468.

Qin, Z. *et al.* (2021) 'Distributed UAV-BSS trajectory optimization for user-level fair communication service with multi-agent deep reinforcement learning', *IEEE Transactions on Vehicular Technology*, 70(12), pp. 12290–12301. doi: 10.1109/TVT.2021.3117792.

Raper, R. *et al.* (2022) 'Sustainability Budgets: A Practical Management and Governance Method for Achieving Goal 13 of the Sustainable Development Goals for AI Development', *Sustainability 2022, Vol. 14, Page 4019*, 14(7), p. 4019. doi: 10.3390/SU14074019.

Rosic, D., Novak, U. and Vukmirovic, S. (2013) 'Role-based access control model supporting regional division in smart grid system', *Proceedings - 5th International Conference on Computational Intelligence, Communication Systems, and Networks, CICSyN 2013*, pp. 197–201. doi: 10.1109/CICSYN.2013.59.

Saheb, Tahereh, Dehghani, M. and Saheb, Tayebeh (2022) 'Artificial intelligence for sustainable energy: A contextual topic modeling and content analysis', *Sustainable Computing: Informatics and Systems*, 35, p. 100699. doi: 10.1016/J.SUSCOM.2022.100699.

Seid, A. M. *et al.* (2021) 'Collaborative Computation Offloading and Resource Allocation in Multi-UAV-Assisted IoT Networks: A Deep Reinforcement Learning Approach', *IEEE Internet of Things Journal*, 8(15), pp. 12203–12218. doi: 10.1109/JIOT.2021.3063188.

Selfridge, O. G. and Franklin, J. A. (1990) 'The perceiving robot: What does it see? What does it do?', pp. 146–151. doi: 10.1109/ISIC.1990.128453.

Senevirathne, P. R. *et al.* (2019) 'Optimal Residential Load Scheduling in Dynamic Tariff Environment', *2019 IEEE 14th International Conference on Industrial and Information Systems: Engineering for Innovations for Industry 4.0, ICIIS 2019 - Proceedings*, pp. 547–552. doi: 10.1109/ICIIS47346.2019.9063296.

Serna Torre, P. and Hidalgo-Gonzalez, P. (2022) 'Decentralized Optimal Power Flow for time-varying network topologies using machine learning', *Electric Power Systems Research*, 212, p. 108575. doi: 10.1016/J.EPSR.2022.108575.

Siddiqi, U. and Lubkeman, D. (1988) 'EXPERT SYSTEM DISPATCHER'S AID FOR DISTRIBUTION FEEDER FAULT DIAGNOSIS.', *Proceedings of the Annual Southeastern Symposium on System Theory*, pp. 519–523. doi: 10.1109/SSST.1988.17105.

Singstock, N. R. *et al.* (2021) 'Machine Learning Guided Synthesis of Multinary Chevrel Phase Chalcogenides', *Journal of the American Chemical Society*, 143(24), pp. 9113–9122. doi: 10.1021/JACS.1C02971/SUPPL_FILE/JA1C02971_SI_001.PDF.

Skowronek, M. *et al.* (2022) 'Inclusive STEAM education in diverse disciplines of sustainable energy and AI', *Energy and AI*, 7, p. 100124. doi: 10.1016/J.EGYAI.2021.100124.

Soret, B. *et al.* (2022) 'Learning, Computing, and Trustworthiness in Intelligent IoT Environments: Performance-Energy Tradeoffs', *IEEE Transactions on Green Communications and Networking*, 6(1), pp. 629–644. doi: 10.1109/TGCN.2021.3138792.

Stamper, R. (1988) 'Pathologies of AI: Responsible use of artificial intelligence in professional work', *AI & SOCIETY 1988 2:1*, 2(1), pp. 3–16. doi: 10.1007/BF01891439.

Stephan, T. *et al.* (2021) 'Energy and spectrum aware unequal clustering with deep learning based primary user classification in cognitive radio sensor networks', *International Journal of Machine Learning and Cybernetics*, 12(11), pp. 3261–3294. doi: 10.1007/S13042-020-01154-Y/TABLES/15.

Stroemich, C. and Thomas, M. (1997) 'Short-term load forecasting system using adaptive logic networks', *Proceedings of the American Power Conference*, 59–1, pp. 161–165.

Sujan Reddy, A. *et al.* (2022) 'Stacking Deep learning and Machine learning models for short-term energy consumption forecasting', *Advanced Engineering Informatics*, 52, p. 101542. doi: 10.1016/J.AEI.2022.101542.

Sun, J. *et al.* (2021) 'Explainable AI enabled conductor galloping predictor design', *2021 IEEE Power and Energy Society Innovative Smart Grid Technologies Conference, ISGT 2021*. doi: 10.1109/ISGT49243.2021.9372239.

Sun, Q. *et al.* (2022) 'Multi-agent energy management optimization for integrated energy systems under the energy and carbon co-trading market', *Applied Energy*, 324, p. 119646. doi: 10.1016/J.APENERGY.2022.119646.

*The engagements of the management of the French transmission system (RTE, R seau de Transport Fran ais] in the matter of quality in providing electricity* (no date).

Trovato, S. A., Imai, M. and Touchton, R. A. (1990) 'Implementation of an on-line expert system in a nuclear power plant. Reactor emergency action level monitor', *American Society of Mechanical Engineers, Dynamic Systems and Control Division (Publication) DSC*, 23, pp. 1–7.

Tsoka, T. *et al.* (2022) 'Explainable artificial intelligence for building energy performance certificate labelling classification', *Journal of Cleaner Production*, 355, p. 131626. doi: 10.1016/J.JCLEPRO.2022.131626.

Vakulchuk, R., Overland, I. and Scholten, D. (2020) 'Renewable energy and geopolitics: A review', *Renewable and Sustainable Energy Reviews*, 122, p. 109547. doi: 10.1016/J.RSER.2019.109547.

Vale, Z. *et al.* (2011) 'MASCEM: Electricity markets simulation with strategic agents', *IEEE Intelligent Systems*, 26(2), pp. 9–17. doi: 10.1109/MIS.2011.3.

Volkova, A. *et al.* (2022) 'Accountability challenges of AI in smart grid services', *e-Energy 2022 - Proceedings of the 2022 13th ACM International Conference on Future Energy Systems*, pp. 597–601. doi: 10.1145/3538637.3539636.

Volodin, V. S. and Tolokonskij, A. O. (2022) 'Application of Machine Learning for Solving Problems of Nuclear Power Plant Operation', *Studies in Computational Intelligence*, 1032 SCI, pp. 585–589. doi: 10.1007/978-3-030-96993-6_65/COVER.

Wang, X. *et al.* (2021) 'Model-Data Integration Driven Based Power System Frequency Response Model', *2021 IEEE International Conference on Artificial Intelligence and Computer Applications, ICAICA 2021*, pp. 107–112. doi: 10.1109/ICAICA52286.2021.9498140.

Wang, Y. *et al.* (2021) 'Transmission network expansion planning considering wind power and load uncertainties based on multi-agent ddqn', *Energies*, 14(19). doi: 10.3390/EN14196073.

Wehenkel, L. *et al.* (1994) 'Decision tree based transient stability method a case study', *IEEE Transactions on Power Systems*, 9(1), pp. 459–469. doi: 10.1109/59.317577.

Wenninger, S., Kaymakci, C. and Wiethe, C. (2022) 'Explainable long-term building energy consumption prediction using QLattice', *Applied Energy*, 308, p. 118300. doi: 10.1016/J.APENERGY.2021.118300.

Wu, J. *et al.* (2022) 'Evaluation, Analysis and Diagnosis for HVDC Transmission System Faults via Knowledge Graph under New Energy Systems Construction: A Critical Review', *Energies*, 15(21), p. 8031. doi: 10.3390/EN15218031.

Xiao, H. *et al.* (2022) 'Vehicle Selection and Resource Optimization for Federated Learning in Vehicular Edge Computing', *IEEE Transactions on Intelligent Transportation Systems*, 23(8), pp. 11073–11087. doi: 10.1109/TITS.2021.3099597.

Xie, Y., Ueda, Y. and Sugiyama, M. (2021) 'A Two-Stage Short-Term Load Forecasting Method Using Long Short-Term Memory and Multilayer Perceptron', *Energies 2021, Vol. 14, Page 5873*, 14(18), p. 5873. doi: 10.3390/EN14185873.

Xu, C. *et al.* (2022) 'Review on Interpretable Machine Learning in Smart Grid', *Energies 2022, Vol. 15, Page 4427*, 15(12), p. 4427. doi: 10.3390/EN15124427.

Xu, C., Li, C. and Zhou, X. (2022) 'Interpretable LSTM Based on Mixture Attention Mechanism for Multi-Step Residential Load Forecasting', *Electronics 2022, Vol. 11, Page 2189*, 11(14), p. 2189. doi: 10.3390/ELECTRONICS11142189.

Yang, Z. *et al.* (2022) 'Trustworthy Federated Learning via Blockchain', *IEEE Internet of Things Journal*, pp. 1–1. doi: 10.48550/arxiv.2209.04418.

Yardley, T. *et al.* (2015) 'Developing a Smart Grid cybersecurity education platform and a preliminary assessment of its first application', *Proceedings - Frontiers in Education Conference, FIE*, 2015-February(February). doi: 10.1109/FIE.2014.7044273.

Yeckle, J. and Tang, B. (2018) 'Detection of electricity theft in customer consumption using outlier detection algorithms', *Proceedings - 2018 1st International Conference on Data Intelligence and Security, ICDIS 2018*, pp. 135–140. doi: 10.1109/ICDIS.2018.00029.

Yigitcanlar, T. *et al.* (2021) 'Responsible Urban Innovation with Local Government Artificial Intelligence (AI): A Conceptual Framework and Research Agenda', *Journal of Open Innovation: Technology, Market, and Complexity*, 7(1), p. 71. doi: 10.3390/joitmc7010071.

Zamponi, M. E. and Barbierato, E. (2022) 'The Dual Role of Artificial Intelligence in Developing Smart Cities', *Smart Cities 2022, Vol. 5, Pages 728-755*, 5(2), pp. 728–755. doi: 10.3390/SMARTCITIES5020038.

Zarazua de Rubens, G. and Noel, L. (2019) 'The non-technical barriers to large scale electricity networks: Analysing the case for the US and EU supergrids', *Energy Policy*, 135, p. 111018. doi: 10.1016/J.ENPOL.2019.111018.

Zhang, C. Y. *et al.* (2022) 'Multi-UAV Trajectory Design and Power Control Based on Deep Reinforcement Learning', *Journal of Communications and Information Networks*, 7(2), pp. 192–201. doi: 10.23919/JCIN.2022.9815202.

Zhang, S. *et al.* (2023) 'Risk Model and Decision Support System of State Grid Operation Management Based on Big Data', *Lecture Notes on Data Engineering and Communications Technologies*, 122, pp. 419–427. doi: 10.1007/978-981-19-3632-6_51/COVER.

Zhang, Z., Xu, C. and Wu, R. (2022) 'Learning-Based Trajectory Design and Time Allocation in UAV-Supported Wireless Powered NOMA-IoT Networks', *2022 IEEE International Conference on Communications Workshops, ICC Workshops 2022*, pp. 1041–1046. doi: 10.1109/ICCWORKSHOPS53468.2022.9814676.

Zhao, J. *et al.* (2021) 'A Survey: New Generation Artificial Intelligence and Its Application in Power System Dispatching and Operation', *5th IEEE Conference on Energy Internet and Energy System Integration: Energy Internet for Carbon Neutrality, EI2 2021*, pp. 3178–3183. doi: 10.1109/EI252483.2021.9713148.

Zhu, Z. *et al.* (2022) 'Module Against Power Consumption Attacks for Trustworthiness of Vehicular AI Chips in Wide Temperature Range', *https://doi.org/10.1142/S0218001422500124*, 36(3). doi: 10.1142/S0218001422500124.